\documentclass[twoside,twocolumn,tightenlines,showpacs,aps,prb]{revtex4}
\usepackage{graphicx}
\usepackage{subfigure} 
\usepackage[colorlinks=true,linkcolor=blue]{hyperref}%
\usepackage[utf8]{inputenc}  \usepackage[brazil]{babel} 

\usepackage{booktabs}		   % Usar todos os recursos do comando tabular
\usepackage{multirow}		   % Para mesclar as célular na tabela
\usepackage{microtype}          % para melhorias de justificação
\usepackage{hhline}             % linhas duplas na tabela

\newcommand{\PC}[1]{\ensuremath{\left(#1\right)}}        % Parenteses Curvos

\begin{document}

\title{Conductance Peak Density in Disordered Graphene Topological Insulators}

\author{Louis G. C. S. Sá$^1$, A. L. R. Barbosa$^2$, and J. G. G. S. Ramos$^1$}

\affiliation{$^1$  Departamento de F\'{\i}sica, Universidade Federal da Para\'iba, 58297-000 Jo\~ao Pessoa, Para\'iba, Brazil\\
$^2$ Departamento de F\'{\i}sica, Universidade Federal Rural de Pernambuco, 52171-900 Recife, Pernambuco, Brazil}

\date{\today}

\begin{abstract}

We investigate the universal properties of quantum transport in graphene nanowires that engender subtle universal conductance fluctuations. We present results for three of the main microscopic models that describe the sublattice of graphene and generate, as we shall show, all the chiral universal symmetries. The results are robust and demonstrate the widely sought sign of chirality even in the regime of many open channels. The fingerprints paves the way to distinguish systems with sublattice symmetry such as topological insulators from ordinary ones by an order of magnitude. The experimental realization requires a single measurement of the chaotic fluctuations of the associated valleytronics conductante. Through the phase coherence length, our theoretical predictions are confirmed with the data from traditional measurements in the literature concerning quantum magnetotransport.

\end{abstract}
\maketitle

\section{Introduction}

The transport phenomena in disordered mesoscopic systems is strongly affected by the wave behavior of the electron [\onlinecite{RevModPhys.69.731,doi:10.1063/1.5026904,HUANG20181,Huang_2020,RevModPhys.89.045005,RevModPhys.80.1355}]. The wave scattering in nanostructures gives rise to the fundamental phenomena of universal conductance fluctuation (UCF), which depends only on the dimensionality and the symmetries of the corresponding coherent state [\onlinecite {Liu_2013,Berezovsky_2010,PhysRevB.35.1039,Huang_2020,10.1038/s42005-017-0001-4}]. An extraordinary characteristic of the UCF is the nature selection of just few ensembles to describe its emerging properties. Despite the complexity of the mesoscopic device, atomic details are irrelevant and the transport properties depend only on fundamental symmetries. According to Random Matrix Theory (RMT) [\onlinecite{RevModPhys.69.731,RevModPhys.89.015005,PhysRevB.86.155118}], there are three ensembles: (1) circular orthogonal ensemble (COE) ($\beta = 1$), if the Hamiltonian supports time-reversal and spin-rotation symmetries, i.e., if none magnetic field is applied $B  = 0$ and the spin-orbit interaction (SOI) is neglected; (2) circular unitary ensemble (CUE) ($\beta = 2$), if time-reversal symmetry is broken by a magnetic field, $B \neq  0$; (3) circular symplectic ensemble (CSE) ($ \beta = 4$), if spin-rotation symmetry is broken, while the time-reversal symmetry is preserved, i.e., the SOI is non-null.

In bulk state, at the thermodynamic limit, the conductance assumes fixed values in the same material. However, in the mesoscopic regime, fluctuations that seem random appear as a function of some field or external energy that vary from sample-to-sample [\onlinecite{PhysRevB.98.155407,10.1038/srep10997,PhysRevB.99.115421,doi:10.1063/1.5031013,10.1038/s41467-019-12560-4}]. Interestingly, these fluctuations are, in fact, chaotic properties categorized through their amplitudes in any of the universal ensembles previously mentioned, namely they depend only on fundamental symmetries of nature. One way to measure the correlation of such chaotic events is to run an average on the ensemble of achievements from various disordered devices [\onlinecite{PhysRevB.98.155407,PhysRevB.99.115421}]. This exhaustive process of make and measure samples provides the important correlation width scale associated with chaos. Several experimental and theoretical results indicate the universality of this scale, which act as a ``chaotic number” [\onlinecite{PhysRevE.92.022904}]. For parametric variations in energy, for instance, this average on samples allows ones to find the electron dwell time as the inverse of the corresponding autocorrelation width [\onlinecite{PhysRevLett.107.176807,PhysRevE.88.010901,PhysRevE.93.012210,10.1038/srep44900}]. The measurements as a function of the external magnetic field, on the other hand, have the phase-coherence length as the associated physical measurable [\onlinecite{10.1038/s42005-017-0001-4}], a relevant parameter of the quantum scattering.

Recent advances in nanotechnology have allowed the production and control of graphene monolayers with carbon atoms distributed in honeycomb lattice [\onlinecite{RevModPhys.81.109,Mucciolo_2010}]. Graphene has received both experimental and theoretical attention due to its special electronic transport properties [\onlinecite{PhysRevB.95.125427,PhysRevB.77.081410,Berezovsky_2010,RevModPhys.81.109,10.1038/srep10997,PhysRevE.99.032118,PhysRevB.99.214446,10.1038/s42005-017-0001-4}]. Another studies demonstrate the existence of universality in graphene beyond the Wigner-Dyson classes previously mentioned [\onlinecite{PhysRevE.94.062214,10.1038/srep10997,doi:10.1063/1.5010973,PhysRevB.93.115120,PhysRevLett.101.016804,PhysRevB.88.245133,PhysRevB.93.125136,PhysRevB.93.085408,PhysRevB.95.075123}]. In the chaotic mesoscopic regime, RMT predicts the existence of ten symmetries classes according to the Cartan's classification [\onlinecite{PhysRevB.86.155118}], with the three of Wigner being the most established. The honeycomb graphene lattice is divided into two sub-lattices, which givie rise to chiral symmetries, allowing the emulation and control of the other Cartan classes in artificial atoms (quantum dots). The Chiral symmetry is an achievement of more general systems also known as topological insulators [\onlinecite{RevModPhys.82.3045,10.1038/srep10997,PhysRevB.99.115421,doi:10.1063/1.5031013,10.1038/s41467-019-12560-4,doi:10.7566/JPSJ.87.034701}] and the phenomenological counterpart, the relativistic chaos [\onlinecite{doi:10.1063/1.5026904,HUANG20181,Huang_2020}]. However, experimental detection of others symmetries is a hard task given that the chirality seems to disappear according the number of open channels (leads widths) increases subtly. For two or more channels, this signal tends to disappear quickly. 

Faced with this scenario, two questions of experimental and theoretical interest naturally arise. The first concerns the extraction of the magnetic correlation width considering the requirement of several experimental designs and, therefore, the synthesis of a very large ensemble of nanowire samples: Is there a measurable capable of extract the correlation width through a single experimental design? And the second one deals with the characteristic values associated with universality in topological insulators such as graphene: Does the autocorrelation width and consequently the phase-coherence length carry peculiar information of topological insulators? In this work, we give a positive answer to both questions. The observable in question is the density of maxima (local maximum per magnetic field interval) already tested in different systems [\onlinecite{PhysRevLett.107.176807,PhysRevE.88.010901,PhysRevE.93.012210,10.1038/srep44900,PhysRevB.98.155407,PhysRevE.92.022904}]. To extend the validity of our result, we also investigate the graphene monolayer in different scenarios and find different numbers associated with the chaos and universality that can be extracted from a single realization. Our results are confirmed by experimental data available in the literature [\onlinecite{PhysRevLett.104.186802,PhysRevLett.110.156601}].

\begin{figure}[!]
\centering
\includegraphics[width=7cm]{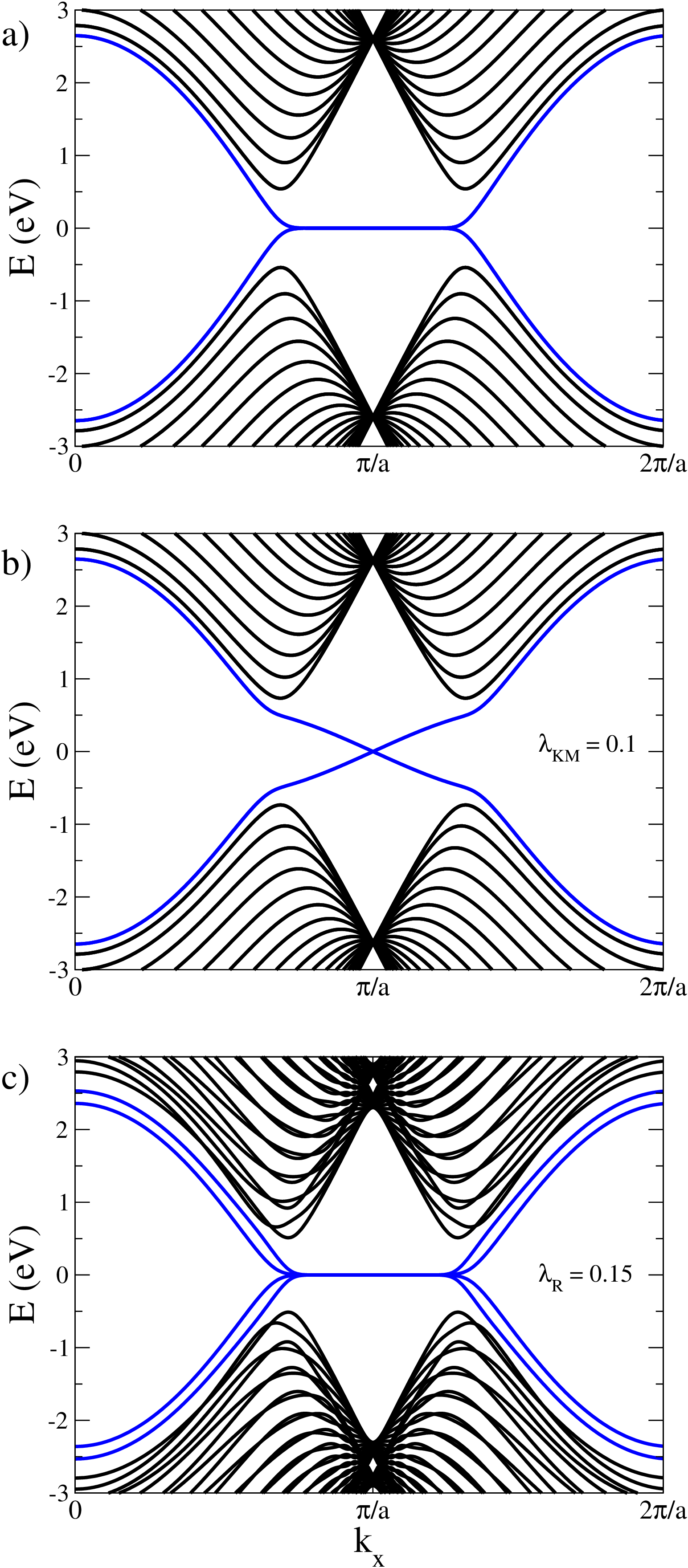}
\caption{Band structures of ZGNR samples with 84 atoms. Blue line indicates the edge states. Band structure of (a) model I, preserved the time-reversal and spin-rotation symmetries, characterizing by COE. The time-reversal symmetry is breaking for (b) model II and preserved for (c) model III, while spin-rotation symmetry is breaking in both models by SOI $\lambda_{KM}$ = 0.1 and $\lambda_R$ = 0.15. The model II and III are characterizing by CUE and CSE, respectively.} 
\label{bands}
\end{figure}

\section{Method}

In this work we investigate the three main models for graphene nanowires. As we shall show, all of them exhibits the UCF. The first model (model I) supports spin-rotation symmetry with neglected SOI terms. On the second (model II) and third (model III) models, we implement the effect of the SOI in the electronic structure, proposed by Kane and Mele [\onlinecite{PhysRevLett.95.226801,PhysRevLett.95.146802}], as a spin-rotation symmetry breaking mechanism. 

Disordered graphene in a tight-binding representation has the Hamiltonian for
honeycomb lattice defined as [\onlinecite{PhysRevLett.95.226801,PhysRevLett.95.146802,10.1038/srep10997,PhysRevB.97.085413}]
\begin{equation}
H_1 = \sum_i \varepsilon_i c_i^\dagger c_i - t \sum_{\langle i,j \rangle} e^{i \phi _{ij}} c_i^\dagger c_j
\end{equation}
for model I,
\begin{equation}
H_2 = H_1 - \frac{i2}{\sqrt{3}} \lambda_{KM} \sum_{\langle \langle i,j \rangle \rangle} e^{i \phi_{ij}} \PC{\hat{\textbf{d}}_{in}  \times \hat{\textbf{d}}_{jn}}_z s_z c_i^\dagger c_j
\end{equation}
for model II and
\begin{equation}
H_3 = H_1 - i \lambda_R \sum_{\langle i,j \rangle} e^{i \phi_{ij}} \PC{\textbf{s} \times \hat{\textbf{d}}_{ij}}_z c_i^\dagger c_j
\end{equation}
for model III, where $\langle ...\rangle$ and $\langle \langle ... \rangle \rangle$ denote the nearest-neighbor and next-nearest-neighbor interactions, respectively. On the model I, the first term introduce short-range disorder with $\varepsilon_i$ randomly chosen in the range ($-W/2<\varepsilon_i<W/2$), being $W$ the measure of the disorder strength and $c_i (c_i^\dagger)$ is the annihilation (creation) operator on the $i$th lattice site. The second term represents an usual nearest-neighbor interaction, with $t$ denoting the hopping between $C$ atoms. Here we choose the value $t$ = 2.6 eV, following DFT calculations [\onlinecite{Choe_2010}]. The time-reversal symmetry breaking is generated by an external magnetic field $B$ accounting the magnetic flux $\phi_{ij} = e/\hbar \int_{\textbf{r}_i}^{\textbf{r}_j} \textbf{A} \cdot d\textbf{l} $. In this work, we use the gauge $\textbf{A} = (-By,0,0)$ as the vector potential for perpendicular magnetic field ($z$-direction) to graphene sheet. The second term contemplated on the model II is the mirror symmetric SOI that involves next nearest sites of indices $i$, $j$ with $n$ being the common nearest neighbor of $i$ and $j$, and, consequently, $\hat{\textbf{d}}_{in}$ describes a vector pointing from $n$ to $i$. The second term on the model III is a nearest neighbor Rashba term. The symbol $\textbf{s}$ denotes the Pauli matrix that describes the electron spin.

We perform tight-binding simulations through the Kwant code [\onlinecite{Groth_2014}]. We calculated the conductance using the Landauer-B\"{u}ttiker formulation, 
$G\ =\ e^2/h \ Tr(tt^\dagger),$
where $t$ is the transmission matrix block of the scattering matrix, written in terms of Green's function. The system is coupled to two semi-infinite ideal leads and the sample-to-sample fluctuation behavior can be characterized by the conductance deviation $\text{rms}[G] =\sqrt{\langle  G^2\rangle -\langle G\rangle^2}$.

\begin{figure}[!]
\centering
\includegraphics[width=8cm]{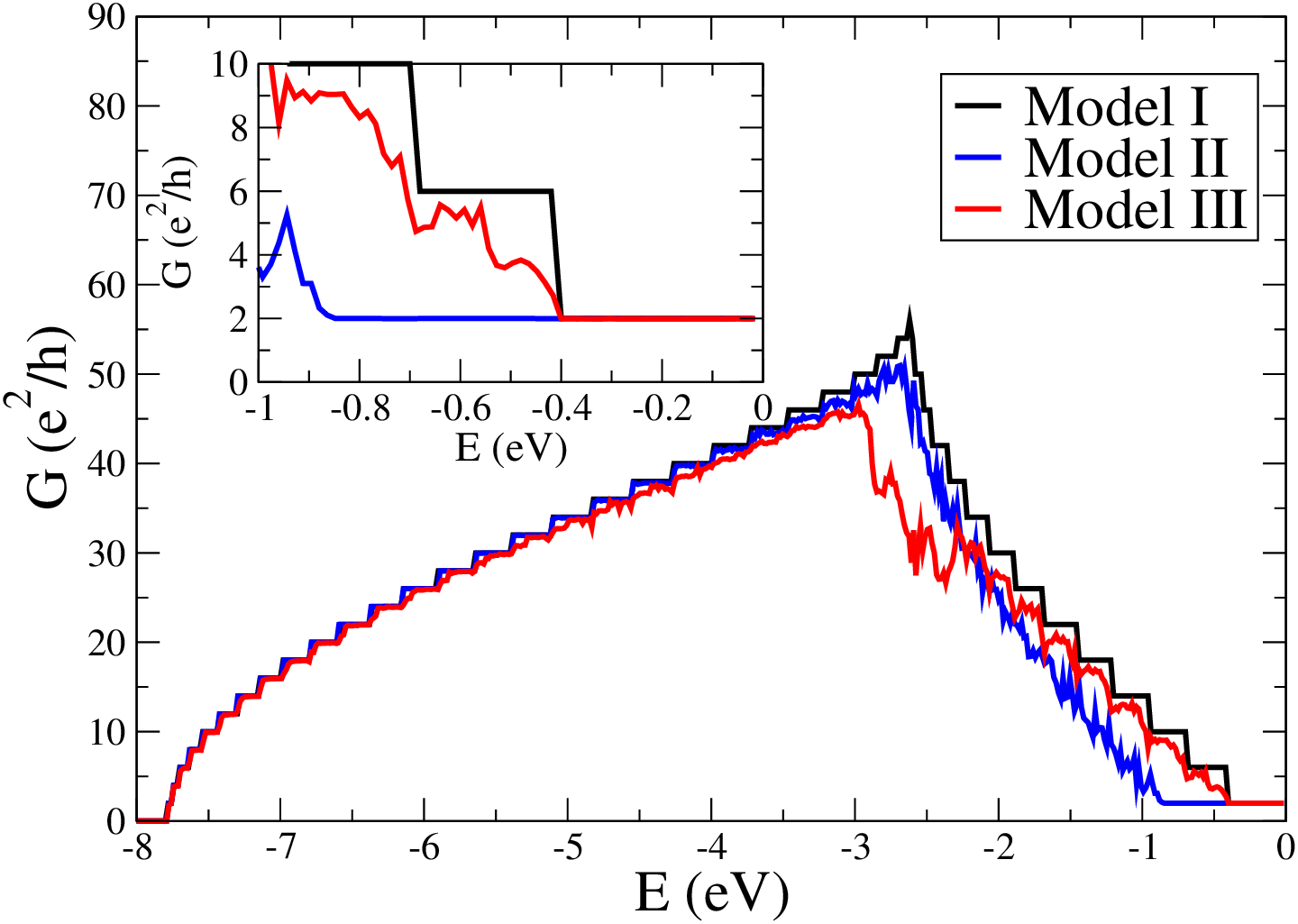}
\caption{Conductance of a graphene nanowire as a function of Fermi energy in absent of both disorder and magnetic field for ZGNR sample with 84 atoms and 100 nm. Inset: the conductance for energy above $-0.4$ eV indicates the edge states are unaffected by the SOI.} 
\label{perfect-Lx150nm}
\end{figure}

\section{Results and Discussions}

For pedagogical reasons, we divide this section in the following four subsections: A) We show the effects of SOI on the graphene band structure and in the corresponding conductance without disorder; B) We incorporate effects of disorder on the graphene conductance and also in its UCF; C) We describe the conductance peak density and analyze the corresponding numerical data using, as a method, results from the principle of maximum entropy; D) The analysis of the conductance peak density will be applied to UCF experimental data from Refs. [\onlinecite{PhysRevLett.104.186802,PhysRevLett.110.156601}].

\subsection{The Graphene Wire in the Absence of Disorder}

We begin the investigation obtaining known results and analyzing the graphene band structure. We explore a zigzag graphene nanribbon (ZGNR) with $84$ atoms, in absence of disorder and magnetic field. The results are depicted in the Fig.(\ref{bands}). For the model I, Fig.(\ref{bands}.a) shows that the bands connected at Fermi energy ($E$=0) are populated by the edge states and the other ones are the bulk bands, unveiling degenerate copies for each band. For the model II, according to the results shown in the Fig(\ref{bands}.b), the effect of the spin-rotation symmetry breaking with the SOI is to open the edge bands and the gap	undergoes an increment of 1.0 eV to 1.5 eV, which is in accordance with the Kane-Mele model [\onlinecite{PhysRevLett.95.226801}]. The latest model provides results explaining that the edge states are not chiral since each edge has propagating states in both directions. The model III contemplates the Rashba term which violates $z \longrightarrow -z$ \ mirror symmetry [\onlinecite{PhysRevLett.108.166606}], shifting some bands as depicted in the Fig.(\ref{bands}.c).

We investigate the ZGNR conductance in a sample with 84 atoms and 100 nm of length, in absence of disorder and magnetic field. Results for the three models are shown in the Fig.(\ref{perfect-Lx150nm}). Without SOI (model I), the conductance is quantized, as expected, and it is null for any energy out of the range $|E|\ >\ 8.0$ eV, while its maximum value is at $|E| = 2.6$ eV. With intrinsic SOI in the scattering region (model II, $\lambda_{KM}=0.1$), the conductance steps show fluctuations, and the edge state decreases from $-0.4$ eV to $-0.8$ eV compared to the model I, as expected by the general behavior of the band structures. The Rashba SOI (model III, $\lambda_{R}=0.15$) also induces conductance fluctuation.

\subsection{Disordered Graphene Wire}

\begin{figure}[!]
\centering
\includegraphics[width=8cm]{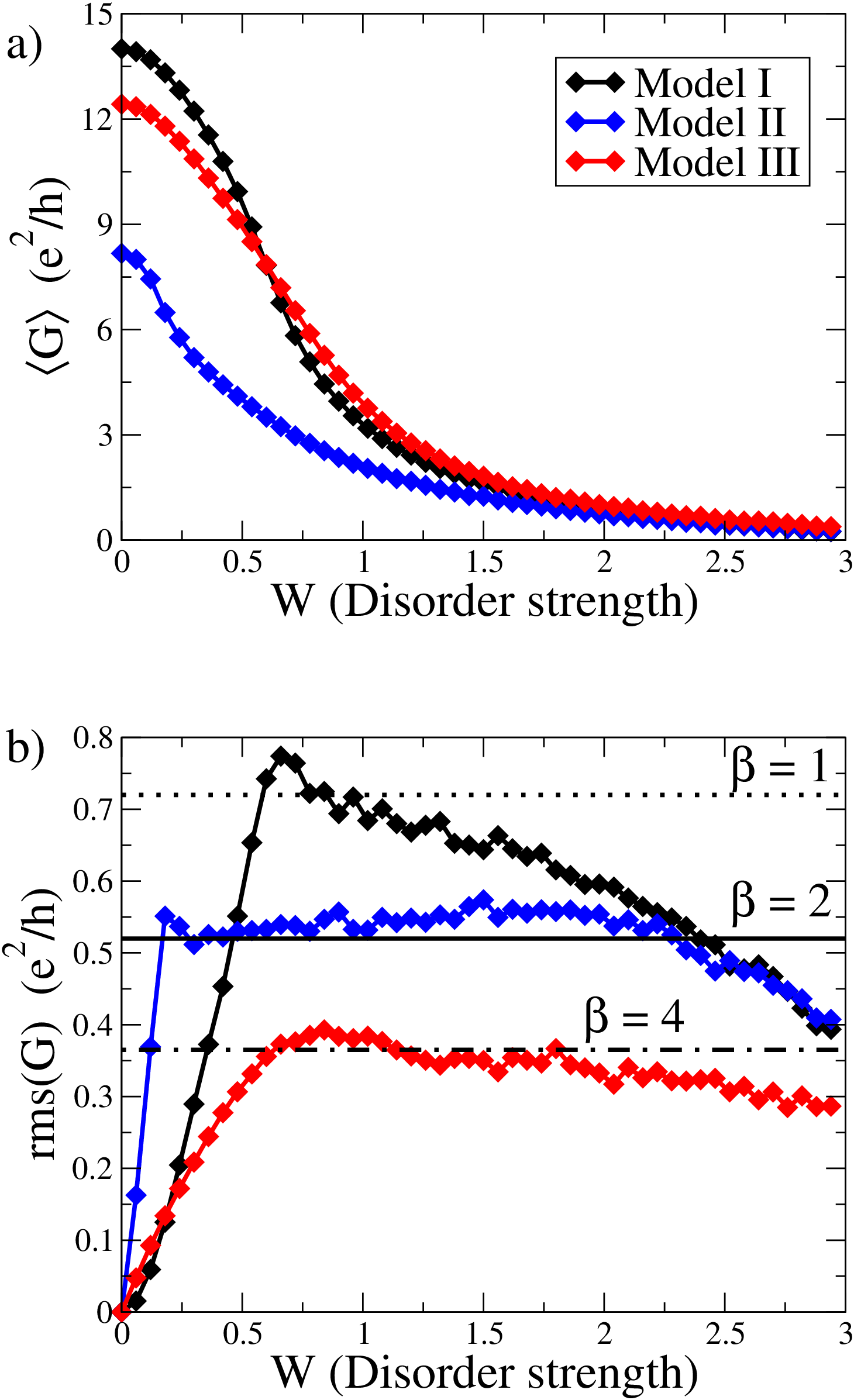}
\caption{Disorder effects in graphene. (a) Conductance average and (b) its deviation as a function of disorder strength $W$ at an energy of $-1.2$ eV. The lines in (b) represent the deviation values predicted by the RMT for COE ($\beta = 1$), CUE ($\beta = 2$), and CSE ($\beta = 4$).} 
\label{graphene_W}
\end{figure}

The main purpose of this present investigation is to simulate samples whose relevant properties are manifest in the UCF whenever the electron transport is diffusive. Within this general purpose, we analyze the conductance average and its deviation as a function of the disorder strength $W$, as shown in the Fig.(\ref{graphene_W}), for which we take the typical values $\lambda_{KM} = 0.15$ and $ \lambda_R = 0.15 $ on models II and III, respectively. Although the absence of disorder, $W=0$, can induce the system to behave as ideal, the conductance decreases according the disorder is magnified, as depicted in the Fig.(\ref{graphene_W}.a), and it is also the disorder that induces the sample-to-sample fluctuation, Fig.(\ref{graphene_W}.b). Therefore, with moderate values of $W$, the diffusive regime is activated, and indicates that the conductance deviations Fig.(\ref{graphene_W}.b) support an expected characteristic value of a quasi-one-dimension nanowire, described in the framework of RMT [\onlinecite{RevModPhys.69.731}]. For large values of $W$, the conductance performs a conductor/insulator transition occasioned by the Anderson localization [\onlinecite{RevModPhys.80.1355}], i.e., the conductance and its deviation tends to zero, as expected.

\begin{figure}[!]
\centering
\includegraphics[width=8cm]{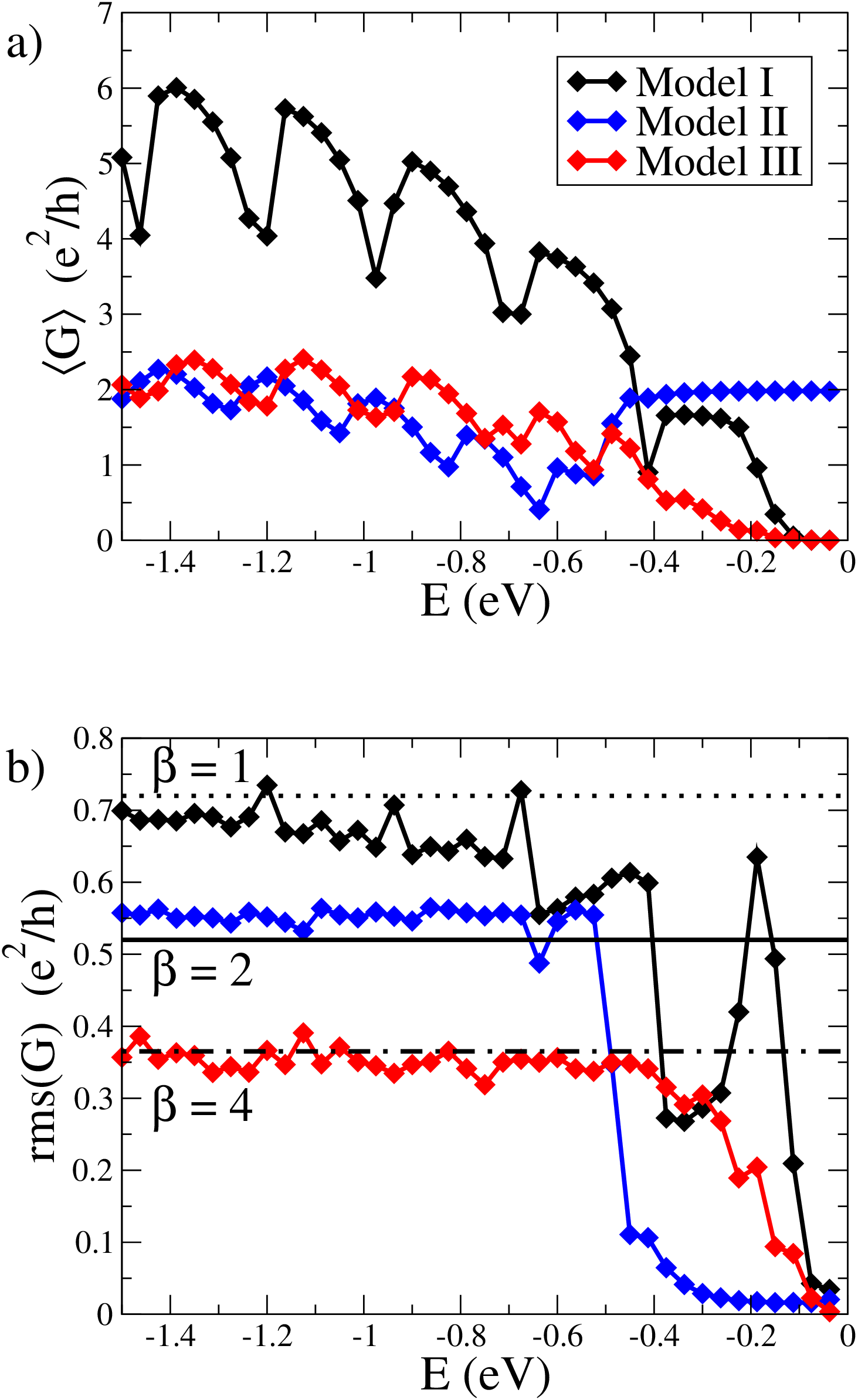}
\caption{(a) Conductance average and (b) its dviation as a function of Fermi energy in presence of disorder $W$ = 0.75. The lines in (b) represent the deviation values predicted by the RMT for COE ($\beta$ = 1), CUE ($\beta$ = 2) and CSE ($\beta$ = 4). } 
\label{energies}
\end{figure}

\begin{figure}[!]
\centering
\includegraphics[width=8cm]{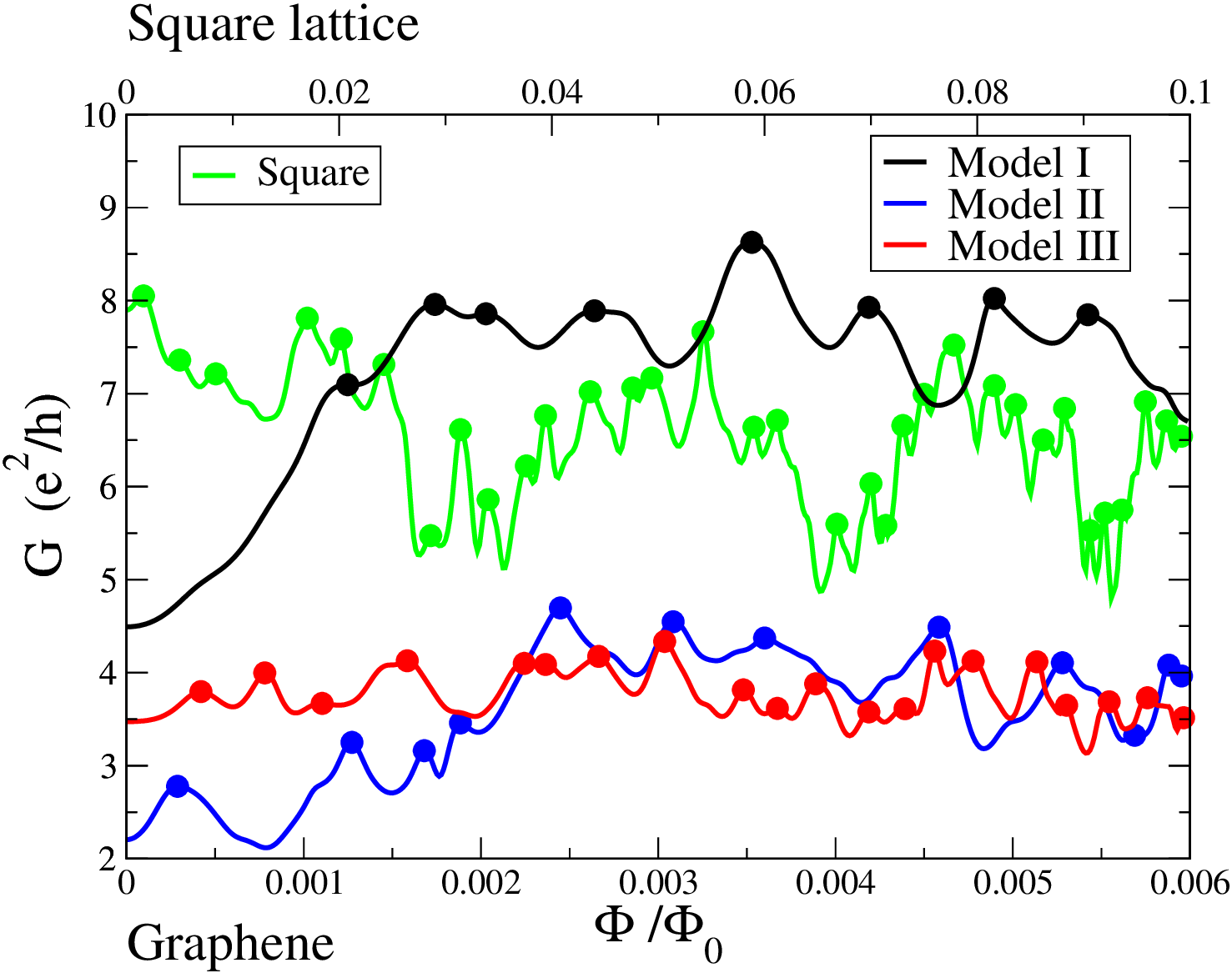}
\caption{Conductance as a function of perpendicular magnetic flux, with a disorder strength of $W$ = 0.75 at an energy of $E$ = -1.2 eV. For model II and III, the values used are $\lambda_{KM}$ = 0.15 and $\lambda_R$ = 0.15, respectively.}
\label{graphene_peaks}
\end{figure}

Another perspective of the UCF can be enlightened through the conductance average and its deviation as a function of the Fermi energy, Fig.(\ref{energies}), with $\lambda_{KM} = 0.15$ and $ \lambda_R = 0.15 $ for models II and III, respectively, and a disorder value $W = 0.75$. As expected, the conductance deviation goes to the COE ($\beta=4$) value 0.74 $e^2/h$ for model I, as depicted in the Fig.(\ref{energies}.b). For model II, Fig.(\ref{energies}.a), the edge states ($E > $ -0.4 eV) are unaffected by disorder and indicates a ballistic behaviour of the electron transport. The robustness of the topological edge states [\onlinecite{RevModPhys.82.3045}], although the conductance fluctuation, Fig.(\ref{energies}.b), exhibits the universal behavior with the UCF value of 0.52 $e^2/h$ in the diffusive regime ($E < $ -0.4 eV), a value that correspond to the CUE ($\beta=2$). Notice that the model II correspond to the CSE ($\beta=4$) in the framework of RMT. As discussed by Choe and Chang [\onlinecite{10.1038/srep10997}], the distinct UCF value in the Kane-Mele model is attributed to the particular form of $H_2$ which can be written as a sum of two Haldane Hamiltonians [\onlinecite{RevModPhys.82.3045}], $H_2 = H^+_{Haldane} \oplus H^-_{Haldane}$, a direct sum of spin-up and spin-down Haldane terms with each component supporting opposite sign. The Haldane model is categorized as the circular unitary ensemble ($\beta$ = 2) since the phase acquired by the next-nearest-neighbor hopping term breaks the time-reversal symmetry. Hence, the model II exhibits a UCF value 0.52 $e^2/h$ whereas its Hamiltonian is founded in the Haldane model. For the model III, the edge states are affected by disorder as indicates the Fig.(\ref{energies}.b) and the conductance deviation converges to the GSE value 0.37 $e^2/h$, as expected.

\subsection{Conductance Peak Density}

In order to investigate the connection between the conductance peak density and its correlation function, we analyze the conductance behavior as a function of a perpendicular magnetic field. Effectively, in all the simulations we use a disorder strength $W = 0.75$ and a Fermi energy tuned in $-1.2$ eV. The application of a perpendicular magnetic field flux in the sample gives rise to a crossover COE-CUE for the model I and a CSE-CUE for the model III, in both cases this is due to the time-reversal symmetry breaking. The model II is unaffected. 

The Fig.(\ref{graphene_peaks}) shows typical curves of conductance, which allows ones to count the maxima number. The Conductance Peak Density (CPD) can be defined as the ratio between the maxima number $N$ and the range of dimensionless perpendicular magnetic flux, $\rho_{\Phi} =N/(\Delta \Phi/\Phi_0)$  [\onlinecite{PhysRevLett.107.176807}]. Hence, we build the central sector of the Table I from the Fig.(\ref{graphene_peaks}), which shows the maxima number and the CPD for the three models. Notice we use in all graphene models the magnetic flux range $\Delta \Phi/\Phi_0 =0.006$ (bottom horizontal axis) and, for the square lattice, the range $\Delta \Phi/\Phi_0 =0.1$ (top horizontal axis) as indicated in the Fig.(\ref{graphene_peaks}).

\begin{figure}[!]
\centering
\includegraphics[width=8cm]{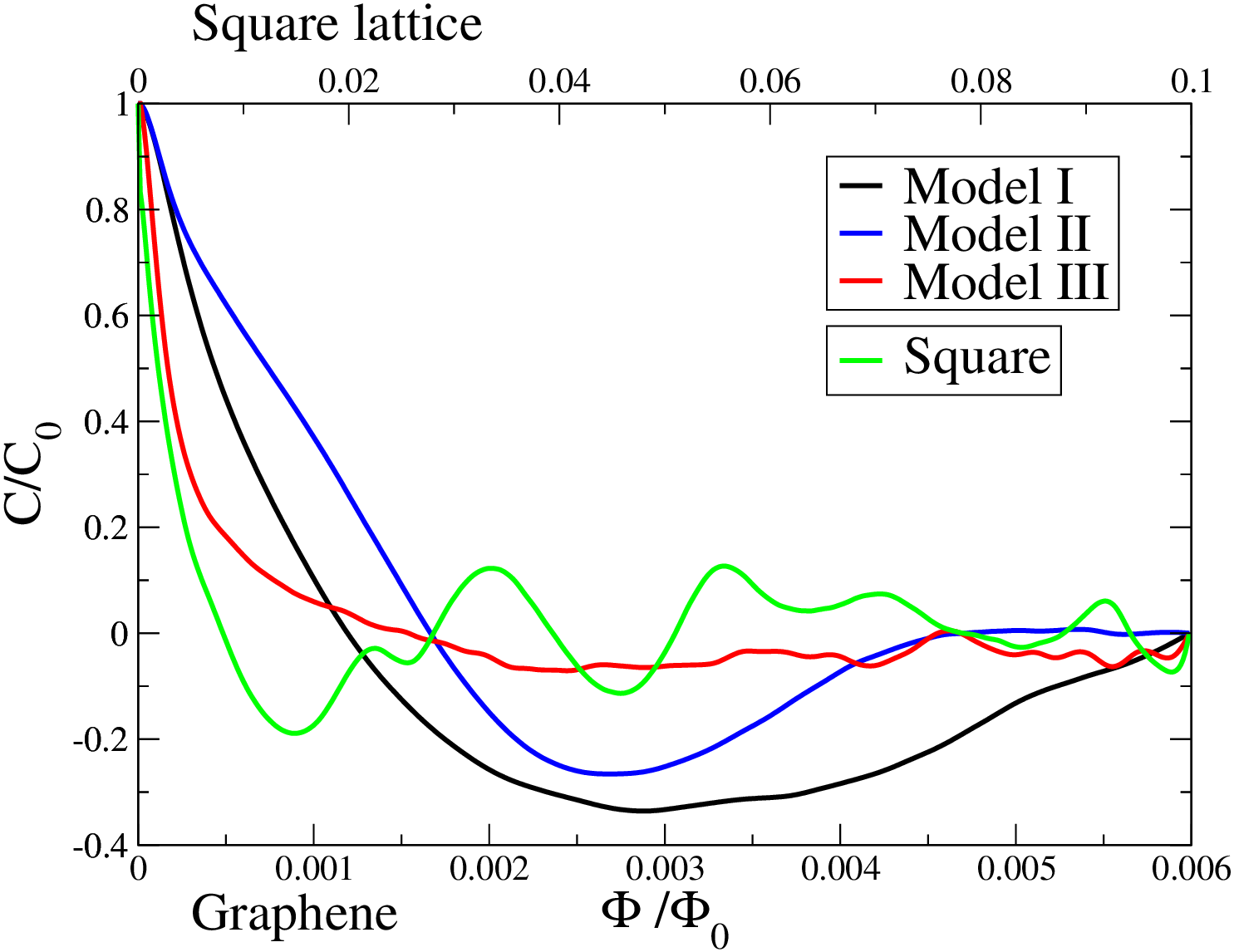}
\caption{Conductance correlation in function of perpendicular magnetic flux obtained from $10^{3}$ realizations. For model II and III were used $\lambda_{KM}$ = 0.15 and $\lambda_R$ = 0.15, respectively.} 
\label{correlation}
\end{figure}

The use of the maximum entropy principle in quantum systems can establish an important connection between the autocorrelation width, $\Gamma_{\Phi}$, and the density of maxima, $\rho_{\Phi}$, as proposed in the reference [\onlinecite{PhysRevLett.107.176807}]. The method was applied in a variety of scenarios, yielding applications on different systems [\onlinecite{PhysRevE.88.010901,PhysRevE.93.012210}]. The experimental obtention of the correlation width requires an average under the data derived from the synthesis and measurement on a ensemble of samples. Therefore, it is a costly procedure, although it results in this important characteristic number of chaos. The requirement of a large amount of data in order to extract the average in the ensemble can be replaced by a simple and unique measurement through the maxima density observable. The method determines such relation through the formula
\begin{equation}
\rho_{\Phi} = \frac{3}{\pi \sqrt{2} \Gamma_{\Phi}} \approx \frac{0.68}{\Gamma_{\Phi}}.
\label{eq-peakdensity}
\end{equation}
To confirm the results obtaining previously, present in the central sector of the Table I, we calculate the autocorrelation function $C(\Delta (\Phi/\Phi_0))$ by simulating several samples through subtly modifications in the boundary conditions of the wire in each sample in order to establish a connection with the chaos. Once with the data, we extract the $\Gamma_{\Phi}$. The former was calculated through its usual definition  
$$
C(\Delta {\Phi/\Phi_0}) = \langle G(\Delta {\Phi/\Phi_0})G(0) \rangle-\langle G(\Delta \Phi/\Phi_0) \rangle\langle G(0) \rangle,
$$
yielding the results displayed in the Fig.(\ref{correlation}) for the graphene models (bottom horizontal axis) and for the square lattice (top horizontal axis). The autocorrelation width is also defined in the usual way, i.e, the $\Delta \Phi/\Phi_0$ value at half height
$$
\frac{C(\Gamma_{\Phi})}{C(0)} =\frac{1}{2}.
$$
The $\Gamma_{\Phi}$ values obtained from Fig.(\ref{correlation}) are presented in the right side column of the Table I. Substituting the  $\Gamma_{\Phi}$ values in the Eqs.(\ref{eq-peakdensity}), we obtain the CPD, which is also presented on the right side of the table \ref{table}. The results of third and firth columns are in great agreement and demonstrate the efficiency of the CPD procedure for a disorder graphene device.

\begin{table}[!]
  \centering
    \begin{tabular}{l|ll|ll}
    \toprule
    \toprule
    		            &  N \ \ \ \ & $\rho_{\Phi} $  \ \ \ \ & $\Gamma_{\Phi}$ \ \ \ \  & $ \rho_{\Phi} $ \\ \hline
    		   Model I   &  8      & 1333            & 4.4 $\times 10^{-4}$    & 1532 \\
              Model II  &  11      & 1833           & 7.4 $\times 10^{-4}$    & 1614  \\
              Model III &  20    & 3333            & 1.8 $\times 10^{-4}$    & 3778\\
              Square &  35    &   371          & 21.2 $\times 10^{-4}$    & 321\\
    \bottomrule
    \end{tabular}%
    \caption{Second column: the number of maximums (N) of the Fig.(\ref{graphene_peaks}); Third column: the conductance peak density obtain from $\rho_{\Phi/\Phi_0} =N/\Delta \Phi/\Phi_0$ with  $\Delta \Phi/\Phi_0 =0.006$; Fourth column: correlation width length($\Gamma_\perp$) obtained from auto-correlation function Fig. (\ref{correlation}); Fifth column: the conductance peak density obtain from Eq. (\ref{eq-peakdensity}). There is a great agreement between the both methods of obtaining the conductance peak density. }
  \label{table}%
\end{table}%

On the one hand, as depicted in the Figs.(\ref{graphene_W}.b)-(\ref{energies}.b), the conductance deviations of disordered graphene nanowire follow the fundamental symmetries of Wigner-Dyson ensembles, that is, they do not provide any information related to the graphene chyral symmetry. On the other hand, the results in the Fig.(\ref{graphene_peaks}) show a significant change in the UCF, leaving clear the fingerprint of chiral symmetry. These changes affect the CPD and can adequately characterize chiral fundamental symmetries in transport measurements. We performed the same simulation for a square lattice for which there is no sublattice thus referring to the usual fundamental symmetries of Wigner-Dyson ensembles. Our results are demonstrated in the Fig.(\ref{correlation}) and also in table I, confirming the result of Ref.[\onlinecite{PhysRevB.98.155407}]. 

As exposed in the Table I, the CPD of disordered graphene nanowire ranges from $1333$ to $3333$, while for typical nanowire the value is $371$. Even more surprisingly, we show that there is a difference of an order of magnitude in all conductance measurements of a topological insulator (honeycomb lattice of graphene) compared to a typical nanowire (square lattice). Therefore, our result demonstrates that UCF clearly carry information about the fundamental symmetry of the sub-lattice structure.

\begin{figure}[!]
\centering
\includegraphics[width=8cm]{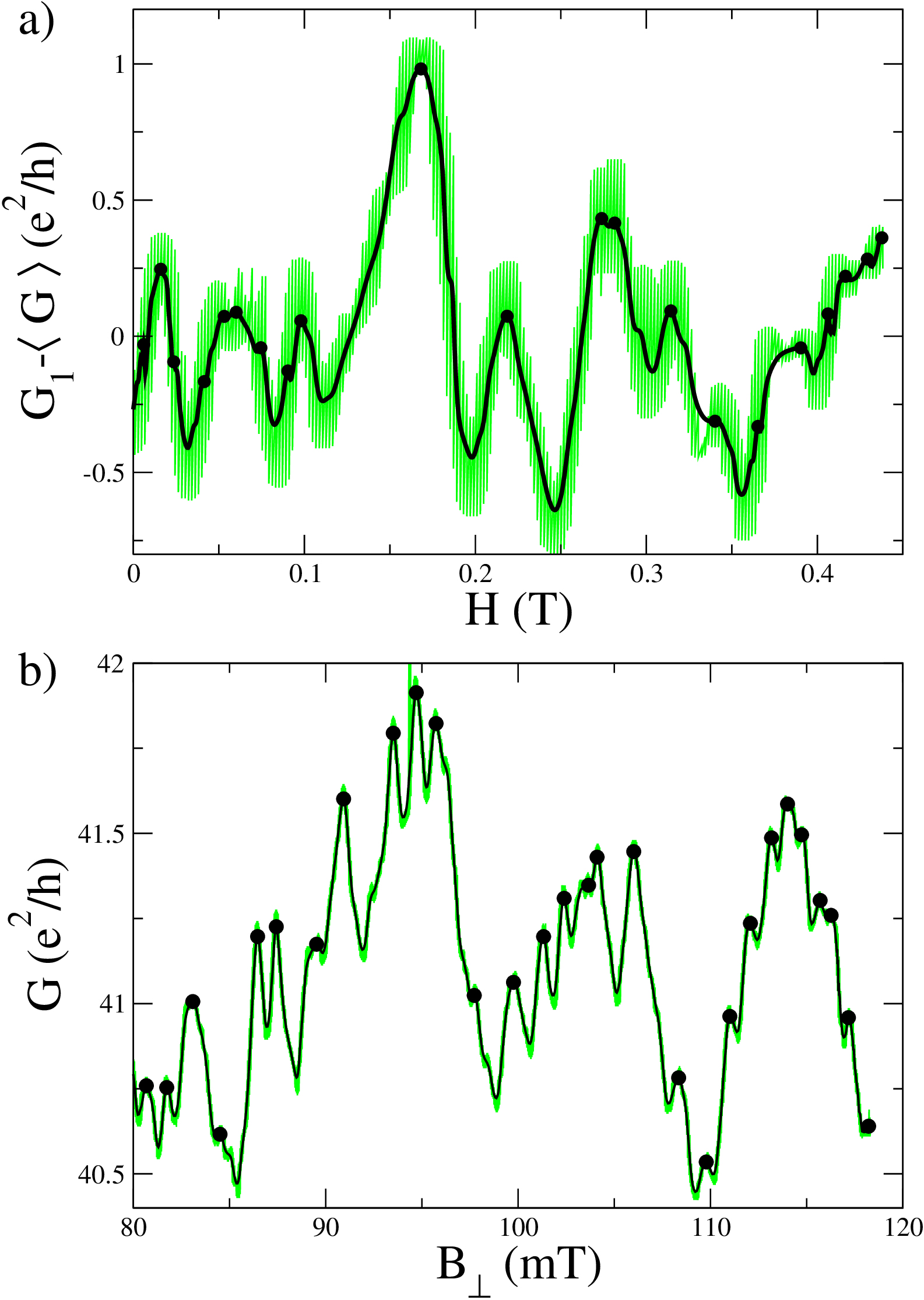}
\caption{Experimental data of a monolayer graphene conductance as a function of the magnetic field. The experimental data (green) is obtained (a) from the Ref.[\onlinecite{PhysRevLett.104.186802}] and (b) from the Ref.[\onlinecite{PhysRevLett.110.156601}]. The smooth of experimental conductance data, in black.} 
\label{experimental}
\end{figure}

\subsection{Chirality Fingerprints Underlying Experimental Signals}

Our results suggest that chirality can be supported even with a high number of open channels, leaving fingerprints on the conductance/UCF. Experimental data on mesoscopic diffusive wires with a number of channels in the order of a few dozen would, consequently, be significantly relevant to prove exactly the autocorrelation width length and other previously established observables. However, the experimental data available is, until our knowledge, for more than $100$ channels. We follow the previous results in order to find, through a single realization, the fingerprints of chirality. Therefore, in this section, we apply the developed methodology in experimental data found in the literature [\onlinecite{PhysRevLett.104.186802,PhysRevLett.110.156601}].  

Ojeda et. al. [\onlinecite{PhysRevLett.104.186802}] and Lundeberg et. al. [\onlinecite{PhysRevLett.110.156601}] developed experimental measures of conductance as a function of the magnetic field in monolayer graphene, whose results are shown in the Fig. (\ref{experimental}. a and b), respectively. In the former, the monolayer graphene nanowire was deposited onto doped silicon and has  dimensions of 2.7 $\mu$m of width and 0.8 $\mu$m of length, while in the latter, it was deposited onto an SiO$_2$/Si wafer and has  dimensional of 4.1 $\mu$m of width and 12.9 $\mu$m of length.
In spite of the experimental sample lengths be one order greater than those used in our numerical simulations, Fig.(\ref{graphene_peaks}), the experimental data have a similar behavior as depicted in the Fig.(\ref{experimental}).

We focus the investigation of the experimental data on one of the most relevant experimental observable, the phase-coherence length $L_{\phi}$, which has a direct relation with the autocorrelation width
\begin{equation}
L_{\phi} = \sqrt{\frac{h}{e\Gamma_\perp}},
\label{pcl}
\end{equation}
with $h$ and $e$ denoting the Planck constant and the electronic charge, respectively. The substitution of the Eq.(\ref{eq-peakdensity}) in the Eq.(\ref{pcl}) render
\begin{equation}
L_{\phi} = \sqrt{\sqrt{2}\frac{\pi h}{3e}\rho_\perp},
\label{Lrho}
\end{equation}
which provides a direct relation between $\rho_\perp$ and $L_{\phi}$. This indicates that we can obtain the phase-coherence length through a simple calculation of the conductance peak density directly from a experimental data even without the information of correlation width.

We first remove the random noise due to both the thermal interference and the experimental apparatus from the data. A simple and straightforward way to perform the extraction is through the B\'ezier algorithm as used and described in the Refs.[\onlinecite{10.1038/srep10997,PhysRevB.98.155407}]. The smooth conductance as a function of a perpendicular magnetic field is depicted by black color in the Fig.(\ref{experimental}). The number of maxima contained in the data of the Fig.(\ref{experimental}.a) is $N=16$ while the magnetic field range is $\Delta B = 0.45$ T, which allows one to directly infer that $\rho_{\perp} = 35.6$ T$^{-1}$. Thereafter, by replacing the value in the Eq.(\ref{Lrho}), we obtain the phase-coherence length $L_{\phi} \approx 0.46$ $\mu$m, which is in agreement with literature of monolayer graphene with mobility $\mu \approx 10^3$ cm$^2$/Vs, Refs.[\onlinecite{PhysRevB.95.125427,doi:10.1063/1.4816721,TERASAWA201714,doi:10.1063/1.4748167,PhysRevLett.108.226602}]. We make a direct comparison between the CPD results of a monolayer graphene and those from the Ref.[\onlinecite{PhysRevB.98.155407}], which investigate InAs nanowire samples (metallic regime). For such metallic samples (Wigner-Dyson ensembles) with mobility also of $\mu \approx 10^3$ cm$^2$/Vs, Ref.[\onlinecite{INAS}], we found the CPD $\rho_{\perp} = 3.4$ T$^{-1}$ and, by replacing the value in the Eq.(\ref{Lrho}), we obtain $L_{\phi} \approx 0.14$ $\mu$m. Remarkable, in similar experimental situations (small mobility values), the CPD of graphene is ten times greater than that of InAs nanowire, which confirm that UCF carry information about the fundamental sublattice symmetry. Also, there is a peculiar coherence length fingerprint in the Universal Chiral Symmetries, confirming nicely our theoretical predictions.

Additionally, the maxima number of the Fig.(\ref{experimental}.b) is $N=32$ and $\Delta B = 0.045$ T, generating the numbers $\rho_{\perp} = 711.1$ T$^{-1}$ and $L_{\phi} \approx 2$ $\mu$m, which is in agreement with literature of monolayer graphene with high mobility $\mu \gg 10^3$ cm$^2$/Vs, Refs.[\onlinecite{Miao1530,PhysRevLett.97.016801,PhysRevLett.102.066801}]. Furthermore, not only we show that the CPD can be understood as an universal sublattice characteristic number but we also obtain a law that relates the coherence phase length with the square root of the maxima density. The rapid oscillation of the conductance as a function of the field in graphene explains its strong quantum coherence behavior.

\section{ Conclusions}

In conclusion, we investigated three widely used models to describe graphene nanowires. The three ones generate universal fluctuations in conductance, each belonging to different classes of fundamental symmetries: orthogonal, unitary or symplectic ensembles. The study demonstrates the connection between the typical spectrum of systems with the sublattice symmetry, the formation of edge states and the classes of universal symmetries. The manifestations in diffusive electron magnetotransport are evident in these different scenarios.

Through the connection between the conductance signals with the principle of maximum entropy, we identified a measurable capable of extracting the correlation length through a single experimental realization of a graphene monolayer. Remarkable, we identified a clear fingerprint of the sublattice structure and, as a deployment, the signal coming from topological insulators by simply counting the maxima number even in the regime of many open channels.

We obtained a law relating the phase coherence length to the square root of the maximum density, $L_{\phi} \propto \sqrt{\rho_\perp}$ Eq. (\ref{Lrho}), showing through experimental data, the sublattice signal by an order of magnitude when compared to the magnetoconductance of usual semiconductor systems. Our study paves the way for the search for coherent quantum transport signals in chiral systems.

\acknowledgments

The work is supported by the Brazilian agencies Coordenação de Aperfeiçoamento de Pessoal de Nível Superior (CAPES), Conselho Nacional de Desenvolvimento Científico e Tecnológico (CNPq), Fundação de Amparo a Ciência e Tecnologia do Estado de Pernambuco (FACEPE), and by Fundação de Apoio à Pesquisa do Estado da Paraíba (FAPESQ-PB).

%\bibliography{ref}

\end{document}